\documentclass[preprint,showpacs,preprintnumbers,amsmath,amssymb]{revtex4}

\usepackage{graphicx}
\usepackage{dcolumn}
\usepackage{bm}
\usepackage{epsfig}
\usepackage{amsfonts}

\begin{document}


\title{Chaotic scalar fields as models for dark energy}

\author{Christian Beck}
\affiliation{
Kavli Institute for Theoretical Physics, University of
California at Santa Barbara, Santa Barbara, CA 93106-4030}
\altaffiliation[Permanent address: ]{
School of Mathematical Sciences, Queen Mary, University of
London, Mile End Road, London E1 4NS, UK}
\email{c.beck@qmul.ac.uk}
\homepage{http://www.maths.qmul.ac.uk/~beck}

\date{\today}

\vspace{2cm}

\begin{abstract}
We consider stochastically quantized self-interacting
scalar fields as suitable models to generate dark energy in the
universe. Second quantization effects lead to new and unexpected
phenomena if the self interaction strength is strong. The
stochastically quantized dynamics can degenerate to a chaotic
dynamics conjugated to a Bernoulli shift in fictitious time, and
the right amount of vacuum energy density can be generated without
fine tuning.
It is numerically observed that the scalar field dynamics
distinguishes fundamental parameters such as the electroweak and
strong coupling constants as corresponding to local minima in the dark
energy landscape. Chaotic fields can offer possible solutions to the
cosmological coincidence problem, as well as to the problem of
uniqueness of vacua.
\end{abstract}

\maketitle











\section{Introduction}

There is by now convincing observational evidence that the
universe is currently in a phase of accelerated expansion
\cite{accel, accel2}. The favored explanation for this behavior is
the existence of vacuum energy or, in a more general setting,
of dark energy. The observations suggest that
the universe currently consist of approximately 73 \% dark energy,
23 \% dark matter, and 4\% ordinary matter \cite{dark}. The nature
and origin of the dominating dark energy component is not
understood, and many different models co-exist. The
simplest models associate dark energy with the vacuum energy of
some unknown self-interacting scalar field, whose potential energy yields
a cosmological constant \cite{cos}. In quintessence models slowly
evolving scalar fields with a nontrivial equation of state are
considered \cite{quin}. String theory also yields possible
candidates of scalar fields who might generate dark energy, in
form of run-away dilatons and moduli fields \cite{string}. Various
exotic forms of matter such as phantom matter \cite{phantom} and
Born-Infeld quantum condensates \cite{BI} are currently being
discussed. For some superstring cosmology ideas related to
small cosmological constants, see also \cite{bafi}.

When trying to formulate a suitable model for dark energy, at
least two unsolved fundamental problems arise:

1. {\em The cosmological constant problem.} Why is the observed
vacuum energy density so small, as compared to typical predictions
of particle physics models? From electroweak symmetry
breaking via the Higgs mechanism one obtains a vacuum energy
density prediction that is too large by a factor $10^{55}$ as
compared to the currently observed value. Spontaneous symmetry
breaking in GUT models is even worse, it yields a discrepancy by a
factor $10^{111}$.

2. {\em The cosmological coincidence problem.} Why is the order of
magnitude of the currently observed vacuum energy density the same
as that of the matter density? A true cosmological constant stays
constant during the expansion of the universe, whereas the matter
energy density decreases with $a^{-3}$, where $a(t)$ is the scale
factor in the Robertson Walker metric. It looks like a very
strange coincidence that right now we live at an epoch where the
vacuum energy density and matter density have the same order of
magnitude, if during the evolution of the universe one is constant
and the other one decreases as $a(t)^{-3}$.

To this list one may add yet another fundamental problem, which we may
call

3. {\em The uniqueness problem.} String theory allows for an enourmous
amount of possible vacua after compactification. In each of these states
the fundamental constants of nature can take on different possible
values. But what is the mechanism that selects out of these infinitely many
possibilities the physically relevant vacuum state, with its associated
fundamental constants that give rise
to a universe of the type we know it
(that ultimately even enabled the development of life)?
Relating the answer purely to an anthropic principle seems unsatisfactory.

In this paper we consider a new model for dark energy which, as
compared to other models, is rather conservative. It just
associates dark energy with self-interacting scalar fields
corresponding to a $\varphi^4$-theory, which is second quantized.
However, the fundamental difference to previous approaches is that
these fields are very strongly (rather than weakly)
self-interacting, and that 2nd quantization effects
play an important role. We
will use as the relevant method to quantize the scalar fields the
stochastic quantization method introduced by Parisi and Wu
\cite{stoch}. In the fictitious time variable of this approach,
the fields will turn out to perform rapid deterministic chaotic
oscillations, due to the fact that we consider not a weakly but a
very strongly self-interacting field. This chaotic behavior is a
new effect not present in any classical treatment. It is generally
well known that chaos plays an important role in general
relativity \cite{chaosgr},
quantum field theories
\cite{chaosqft,book,physicad}, and string theories
\cite{chaosstring}.
The main result of our consideration is that the
chaotic field theories considered naturally
generate a small cosmological constant and have the scope to
offer simultaneous solutions to the
cosmological coincidence and uniqueness problem.

Our physical interpretation is to associate the chaotic behaviour
of the scalar fields with tiny vacuum fluctuations which are allowed
within the bounds set by the uncertainty relation, due to the
finite age of the universe. This
interpretation naturally leads to the right amount of dark energy
density being generated, and fine tuning can be avoided. The
chaotic fields (presently) have a classical equation of state close to $w=-1$,
and can thus account for the accelerated expansion of the
universe. However,
during the early evolution of the universe they behave in a different way:
They effectively track radiation and matter.
This property will
help to avoid the cosmological coincidence
problem.

The chaotic model also contains an interesting symmetry between
gravitational and gauge couplings.
In our model the role of a metric for the 5th coordinate (the fictitious time)
is taken over by dimensionless coupling constants which are given by
the ratio of the fictious time lattice constant and physical time
lattice constant squared (both lattice constants can
still go to zero, just their ratio is fixed). These coupling
constants do not occur in
any classical treatment but are entirely a consequence of our
second quantized treatment.
The vacuum energy generated depends on these couplings in a
non-trivial way.
The physical significance of our model is
illustrated by the fact that we numerically observe the vacuum
energy to have local minima for
coupling constants
that numerically coincide with running electroweak coupling
strengths, evaluated at the known fermionic mass scales, as well
as running strong coupling constants evaluated at the known
bosonic mass scales. This numerical observation,
previously reported in \cite{physicad}, is now
embedded into a cosmological context. The role of the chaotic fields
in the universe can be understood in the sense that they
are responsible for fixing and stabilizing fundamental parameters
as local minima in the dark energy landscape. This is somewhat
similar to the role the dilaton field plays in
string theory after supersymmetry breaking.

Our numerical discovery of local minima that
coincide with known standard model coupling constants makes it very
unlikely that there are different universes
with different fundamental parameters. In fact, the numerical
results provide strong evidence that there is
a unique vacuum state of the universe that
possesses minimum vacuum energy precisely for the known set
of standard model parameters.


This paper is organized as follows. In section 2 we show how a
second-quantized scalar field dynamics can degenerate to a chaotic
dynamics in fictitious time.
Our main example is a chaotic
$\varphi^4$-theory leading to 3rd order Tchebyscheff maps, which
is dealt with in section 3. In section 4 we present a physical
interpretation of the chaotic dynamics using the uncertainty
relation, which in a natural way fixes the order of magnitude of
the vacuum energy density to be generated. Section 5 deals with
energy, pressure and classical equation of state of the chaotic
fields. In section 6 we consider the Einstein equations associated with our
model and discuss a possible way to avoid the cosmological
coincidence problem. Section 7 yields a prediction for the current
ratio of matter energy density and dark energy density to
the critical energy density.
In section 8 we
describe how local minima of the dark energy landscape
generated by the chaotic fields can fix the fundamental
parameters. Finally, in section 9 we discuss spontaneous symmetry breaking
phenomena for the chaotic fields.


\section{Stochastic quantization of strongly self-inter\-acting scalar fields}

Let us consider a self-interacting scalar field $\varphi$ in
Robertson-Walker metric. For a complete theory describing all
quantum mechanical fluctuations we need to second-quantize it.
This can be done via stochastic quantization. In the Parisi-Wu
approach of stochastic quantization one considers a stochastic
differential equation evolving in a fictitious time variable $s$,
the drift term being given by the classical field equation
\cite{stoch}. Quantum mechanical expectations correspond to
expectations with respect to the generated stochastic processes in
the limit $s\to \infty$. The fictitious time $s$ is different from
the physical time $t$, it is just a helpful fifth coordinate to do
2nd quantization. Neglecting spatial gradients the field $\varphi$
is a function of physical time $t$ and fictitious time $s$. The
2nd quantized equation of motion is
\begin{equation}
\frac{\partial}{\partial s}\varphi =\ddot{\varphi}
+3H\dot{\varphi} +V'(\varphi) +L(s,t), \label{sto}
\end{equation}
where $H$ is the Hubble parameter, $V$ is the potential under
consideration and $L(s,t)$ is Gaussian white noise,
$\delta$-correlated both in $s$ and $t$. For e.g.\ a numerical
simulation we may discretize eq.~(\ref{sto}) using
\begin{eqnarray}
s &=& n\tau \\ t &=& i \delta ,
\end{eqnarray}
where $n$ and $i$ are integers and $\tau$ is a fictitious time
lattice constant, $\delta$ is a physical time lattice constant.
The continuum limit requires $\tau \to 0$, $\delta \to 0$, but we
will later argue that it makes physical sense to keep small but
finite lattice constants of the order of the Planck length. We
obtain
\begin{equation}
\frac{\varphi_{n+1}^i-\varphi_n^i}{\tau} = \frac{1}{\delta^2}
(\varphi_n^{i+1}-2\varphi_n^i+\varphi_n^{i-1}) +3\frac{H}{\delta}
(\varphi_n^i-\varphi_n^{i-1}) +V'(\varphi_n^i) + noise
\end{equation}
This can be written as the following recurrence relation for the
field $\varphi_n^i$
\begin{equation}
\varphi_{n+1}^i= (1-\alpha)\left\{ \varphi_n^i
+\frac{\tau}{1-\alpha}
V'(\varphi_n^i)\right\}+3\frac{H\tau}{\delta} (\varphi_n^i-
\varphi_n^{i-1})+\frac{\alpha}{2}(\varphi_n^{i+1}+\varphi_n^{i-1})
+ \tau\cdot noise,
\end{equation}
where a dimensionless coupling constant $\alpha$ is introduced as
\begin{equation}
\alpha:=\frac{2\tau}{\delta^2}.
\end{equation}
We also introduce a dimensionless field variable $\Phi_n^i$ by
writing $\varphi_n^i=\Phi_n^i p_{max}$, where $p_{max}$ is some
(so far) arbitrary energy scale. The above scalar field dynamics
is equivalent to a spatially extended dynamical system (a coupled
map lattice \cite{CML}) of the form
\begin{equation}
\Phi_{n+1}^i=(1-\alpha)T(\Phi_n^i)+\frac{3}{2}H\delta \alpha
(\Phi_n^i-\Phi_n^{i-1})+\frac{\alpha}{2}(\Phi_n^{i+1}+\Phi_n^{i-1})
+ \tau\cdot noise, \label{dyn}
\end{equation}
where the local map $T$ is given by
\begin{equation}
T(\Phi )=\Phi
+\frac{\tau}{p_{max}(1-\alpha)}V'(p_{max}\Phi).\label{map}
\end{equation}
Here the prime means
\begin{equation}
'=\frac{\partial}{\partial
\varphi}=\frac{1}{p_{max}}\frac{\partial}{\partial \Phi}.
\end{equation}
Note that a symmetric diffusively coupled map lattice of the form
\begin{equation}
\Phi_{n+1}^i=(1-\alpha)T(\Phi_n^i)+\frac{\alpha}{2}(\Phi_n^{i+1}+\Phi_n^{i-1})
+\tau \cdot noise \label{sym}
\end{equation}
is obtained if $H\delta << 1$, equivalent to
\begin{equation}
\delta << H^{-1},
\end{equation}
meaning that the physical time lattice constant $\delta$ is much
smaller than the age of the universe. In this case the term
proportional to $H$ in eq.~(\ref{dyn}) can be neglected. The local
map $T$ depends on the potential under consideration. Since we
restrict ourselves to real scalar fields $\varphi$, $T$ is a
1-dimensional map.

The main result of our consideration is that iteration of a
coupled map lattice of the form (\ref{sym}) with a given map $T$
has physical meaning: It means that one is considering the
second-quantized dynamics of a self-interacting real scalar field
$\varphi$ with a force $V'$ given by
\begin{equation}
V'(\varphi)=\frac{1-\alpha}{\tau} \left\{ -\varphi +p_{max}
T\left(\frac{\varphi}{p_{max}}\right) \right\}.
\end{equation}
Integration yields
\begin{equation}
V(\varphi) =\frac{1-\alpha}{\tau} \left\{ -\frac{1}{2} \varphi^2 +
p_{max}\int d\varphi \,T\left(\frac{\varphi}{p_{max}}\right)
\right\} + const .\label{pot}
\end{equation}
In terms of the dimensionless field $\Phi$ this can be written as
\begin{equation}
V(\varphi)=\frac{1-\alpha}{\tau} p_{max}^2 \left\{ -\frac{1}{2}
\Phi^2+\int d\Phi T(\Phi) \right\} +const.
\end{equation}

\section{Chaotic $\varphi^4$-theory}

An interesting observation is the following one. The lattice
constant $\tau$ of fictitious time should be small, in order to
approximate the continuum theory, which is ordinary quantum field
theory. If $\tau$ is small one naively expects the map $T$ given
by eq.~(\ref{map}) to be close to the identity for finite forces
$V'$, since $\tau V'/p_{max}$ is small. What about, however, very
strong forces $V'$ due to very strongly self-interacting fields?
If $p_{max}/\tau$ is of the same order of magnitude as $V'$ then a
nontrivial map $T$ can arise. In particular, this map may even
exhibit chaotic behaviour.

As a distinguished example of a $\varphi^4$-theory generating
strongest possible chaotic behaviour, let us consider the map
\begin{equation}
\Phi_{n+1}=T_{-3}(\Phi_n)=-4\Phi_n^3+3\Phi_n
\end{equation}
on the interval $\Phi\in [-1,1]$. $T_{-3}$ is the negative
third-order Tchebyscheff map, a standard example of a map
exhibiting strongly chaotic behaviour. It is conjugated to a
Bernoulli shift, thus generating the strongest possible stochastic
behavior possible for a smooth low-dimensional deterministic
dynamical system. The corresponding potential is given by
\begin{equation}
V_{-3}(\varphi)=\frac{1-\alpha}{\tau}\left\{
\varphi^2-\frac{1}{p_{max}^2} \varphi^4\right\}+const, \label{16}
\end{equation}
or, in terms of the dimensionless field $\Phi$,
\begin{equation}
V_{-3}(\varphi)=\frac{1-\alpha}{\tau} p_{max}^2 ( \Phi^2 -\Phi^4)
+ const. \label{17}
\end{equation}
Apparently, starting from this potential we obtain
by second quantization a
field $\varphi$ that
rapidly fluctuates in fictitious time on some finite interval,
provided that initially $\varphi_0\in [-p_{max},p_{max}]$.
The small noise term in eq.~(\ref{sym}) can be neglected as
compared to the deterministic chaotic fluctuations of the field.

Of physical relevance are the expectations of suitable observables
with respect to the ergodic chaotic dynamics. For example, the
expectation $\langle V_{-3} (\varphi )\rangle$ of the potential is
a possible candidate for vacuum energy in our universe. One
obtains
\begin{equation}
\langle V_{-3}(\varphi)\rangle =\frac{1-\alpha}{\tau} p_{max}^2 (
\langle \Phi^2\rangle  -\langle \Phi^4\rangle ) + const.
\end{equation}
For uncoupled Tchebyscheff maps ($\alpha=0$), expectations of any
observable $A$ can be evaluated as the ergodic average
\begin{equation}
\langle A \rangle = \int_{-1}^{+1} A(\Phi) d\mu (\Phi),
\label{obs}
\end{equation}
with the natural invariant measure being given by
\begin{equation}
d\mu (\Phi) =\frac{d\Phi}{\pi\sqrt{1-\Phi^2}} \label{mu}
\end{equation}
(see any textbook on chaotic dynamics, e.g. \cite{escort}). This
measure describes the probability distribution of the iterates
under long-term iteration. From eq.~(\ref{mu}) one obtains
$\langle \Phi^2 \rangle =\frac{1}{2}$ and $\langle \Phi^4 \rangle
=\frac{3}{8}$, thus
\begin{equation} \langle V_{-3} (\varphi)
\rangle=\frac{1}{8}\frac{p_{max}^2}{\tau} +const.
\end{equation}

Alternatively, we may consider the positive Tchebyscheff map $T_3
(\Phi)=4\Phi^3-3\Phi$. This basically exhibits the same dynamics
as $T_{-3}$, up to a sign. Repeating the same calculation we
obtain
\begin{equation}
V_{3}(\varphi)=\frac{1-\alpha}{\tau}\left\{
-2\varphi^2+\frac{1}{p_{max}^2} \varphi^4\right\}+const \label{22}
\end{equation}
and
\begin{equation}
V_{3}(\varphi)=\frac{1-\alpha}{\tau} p_{max}^2 ( -2\Phi^2
+\Phi^4). \label{23}
\end{equation}
For the expectation of the vacuum energy one gets
\begin{equation}
\langle V_{3}(\varphi)\rangle =\frac{1-\alpha}{\tau} p_{max}^2 (
-2\langle \Phi^2\rangle  +\langle \Phi^4\rangle ) + const,
\end{equation}
which for $\alpha=0$ reduces to
\begin{equation}
\langle V_{3} (\varphi) \rangle=-\frac{5}{8}\frac{p_{max}^2}{\tau}
+const.
\end{equation}
Symmetry considerations between $T_{-3}$ and $T_3$ suggest to take
the additive constant $const$ as
\begin{equation}
const=+\frac{1-\alpha}{\tau} p_{max}^2 \frac{1}{2} \langle \Phi^2
\rangle.
\end{equation}
In this case one obtains the fully symmetric equation
\begin{equation}
\langle V_{\pm 3} (\varphi)\rangle =\pm
\frac{1-\alpha}{\tau}p_{max}^2 \left\{ -\frac{3}{2} \langle \Phi^2
\rangle +\langle \Phi^4 \rangle \right\},
 \label{symme}
\end{equation}
which for $\alpha\to 0$ reduces to
\begin{equation}
\langle V_{\pm 3} (\varphi)\rangle =\pm \frac{p_{max}^2}{\tau}
\left( -\frac{3}{8} \right).
 \label{symmetry}
\end{equation}


The simplest model for dark energy in the universe, as generated by
a chaotic $\varphi^4$-theory, would be to identify
$\frac{3}{8}p_{max}^2/\tau =\rho_\Lambda$, the constant vacuum
energy density corresponding to a classical cosmological constant
$\Lambda$, which stays constant during the expansion of the
universe. This is certainly a possible simple model. On the other
hand, such an approach would neither solve the cosmological
constant nor the cosmological coincidence problem. For this reason
we will turn to a more advanced model in the following sections, which will
naturally
produces the right amount of vacuum energy density in the
universe.

Before proceeding to this model, let us provide some general comments on
the physical meaning of the parameter $\tau$. The vacuum energy
generated by the chaotic fields is inversely proportional to $\tau$ (see 
eq.~(\ref{symmetry})). 
Strict equivalence of the stochastic quantization method with quantum
field theory requires the continuum limit $\tau \to 0$. In this limit
eq.~(\ref{symmetry}) can still generate a finite amount of vacuum energy,
provided both $\tau \to 0$ and $p_{max}\to 0$ such that
the ratio $p_{max}^2/\tau$ stays finite. 
In fact,
many of the results presented in this paper depend
only on the ratio $p_{max}^2/\tau$ and not on
the individual values attributed to $\tau$ and $p_{max}$.

From the viewpoint of ordinary
quantum field theory, the limit $\tau \to 0$ and $p_{max}\to 0$
means that one formally considers
a potential $V(\varphi)=\mu^2 \varphi^2 + \lambda \varphi^4$
for which both potential parameters
$\mu^2 \sim \tau^{-1}$ and $\lambda \sim \tau^{-1} p_{max}^{-2}$
diverge (see eq.~(\ref{22})), moreover the field $\varphi$ lives on an
infinitely small support $[-p_{max},p_{max}]$. Clearly,
this is a very singular type of quantum field theory, which in principle
can also be studied by other means than stochastic quantization,
though perturbation theory will not be applicable. The advantage
of our formulation in terms of stochastic quantization is that for
the dimensionless chaotically evolving field $\Phi$ the potential
remains finite (see, e.g., eq.~(\ref{23})).
If there is no fictitious time,
then the parameter $\tau$
enters in form of the
(singular) potential parameters $\mu^2\sim \tau^{-1}$
and $\lambda \sim \tau^{-1}p_{max}^{-2}$. 
In the next section we will argue that on physical grounds it makes sense
to consider very small but finite $\tau$. 


\section{Reproducing the currently measured dark energy density}



To obtain quantitative statements on the dark energy density
as generated
by some chaotically evolving field $\varphi$,
let us fix the free parameters
$\tau$ and $p_{max}$ by some physical arguments. Let us start with
the parameter $\tau$. It is the lattice constant of fictitious
time $s$ and has dimension $GeV^{-2}$. Ordinary stochastic
quantization based on Gaussian white noise requires the continuum
limit $\tau \to 0$. But since quantum field theory runs into
difficulties at the Planck scale $m_{Pl}$ and is expected to be
replaced by a more advanced theory at this scale, it is most
reasonable to take the small but finite value
\begin{equation}
\tau \sim m_{Pl}^{-2}. \label{tau}
\end{equation}

Next, consider the parameter $p_{max}$. It has dimension $GeV$ and
describes the maximum energy scale of our rapidly fluctuating
scalar fields $\varphi$, who take on values on the finite interval
$[-p_{max},p_{max}]$. A natural value of $p_{max}$ follows if one
associates the rapidly fluctuating chaotic fields $\varphi_n^i$
with vacuum fluctuations that are allowed due to the uncertainty
relation
\begin{equation}
\Delta E \Delta t=O(\hbar) \label{H}.
\end{equation}
Taking $\Delta t\sim t $ of the order of the age of the universe,
a corresponding energy uncertainty $\Delta E$ arises. This $\Delta
E$ is very large for a very young universe, and then decreases to
extremely small values for the current age of the universe of
about 13.7 Gyr. Any finite age $t$ of the universe implies that
spontaneous vacuum fluctuations with energies of order $\Delta
E\sim t^{-1}$ can occur. It is physically plausible to identify
these energy fluctuations $\Delta E$ with the rapidly fluctuating
chaotic fields $\varphi =p_{max}\Phi_n^i$, since both $\Delta
E$ and $\varphi$ live on a finite interval, and both fluctuate in
an unpredictable way.
The uncertainty relation (\ref{H}) together with $\Delta t \sim t$
implies (in units where $\hbar =c=1$)
\begin{equation}
p_{max}\sim \frac{1}{t}. \label{HH}
\end{equation}
In this way the energy scale $p_{max}$ occurring in the chaotic
field theories is most naturally identified with the inverse age
of the universe.
However, note that
this quantum mechanical interpretation in terms of vacuum
fluctuations requires a {\em chaotic} map $T$, since some
regularly evolving $\Phi_n^i$ cannot be associated with
fluctuations at all.

It is remarkable that by taking eq.~(\ref{tau}) and eq.~(\ref{HH})
together, the right amount of vacuum energy
follows without any fine tuning. One has for generic
chaotic maps $T$
\begin{equation}
\langle V(\varphi)\rangle \sim \frac{p_{max}^2}{\tau} \sim H^2
m_{Pl}^2.\label{33}
\end{equation}
since $t^{-1}\sim H$. Moreover,
\begin{equation}
H^2=\frac{8\pi G}{3} \rho_c \sim \frac{1}{m_{Pl}^2}\rho_c
\label{34}
\end{equation}
where $\rho_c$ denotes the critical density of a flat universe and
$G=m_{Pl}^{-2}$ is the gravitational constant. Combining
eq.~(\ref{33}) and (\ref{34}) one obtains
\begin{equation}
\langle V(\varphi) \rangle \sim \rho_c,\label{rhoc}
\end{equation}
as required and confirmed by current astronomical observations.
Our simple physical interpretation, namely to interpret the
chaotic fluctuations as vacuum fluctuations allowed due to the
finite age of the universe, thus yields the right order of
magnitude of dark energy density.



In general, eq.~(\ref{rhoc}) only yields order-of magnitude
estimates.
Nevertheless, it is instructive to work out some
concrete numbers, based on simple model assumptions. For example,
we may assume that
the entire vacuum energy of the universe is due to
one chaotic field described by $V_{-3}$.
The current age of the
universe is $t_0=(13.7\pm 0.2)$ Gyr $=(4.32\pm 0.06)\cdot
10^{17}s$ \cite{dark}. Using an uncertainty relation of the form
$\Delta E \Delta t=\hbar /2$ we get $p_{max}=1/(2t_0)=(7.62 \pm
0.08)\cdot 10^{-43}$ GeV. Choosing $\tau= \kappa m_{Pl}^{-2}$,
where $\kappa$ is some dimensionless number of $O(1)$, we get
\begin{equation}
\langle V(\varphi) \rangle =\frac{3}{8}\frac{p_{max}^2}{\tau} =(3.19\pm
0.05)\cdot 10^{-47} \kappa^{-1}\;GeV^4 \label{35}
\end{equation}
The current observational estimate of dark energy density in the
universe is \cite{dark}
\begin{equation}
\rho_\varphi^{Obs} =(2.9\pm 0.3)\cdot 10^{-47} GeV^4,
\end{equation}
which is consistent with $\kappa \approx 1$. If the observed dark
energy in the universe is produced by our chaotic theory, then the
measured data imply
\begin{equation}
\kappa =1.10\pm 0.10.
\end{equation}
Once again, this estimate is based on concrete model assumption.
In general, we do do not know the precise
values of the proportionality constants in eq.~(\ref{tau}) and (\ref{HH}),
moreover we do not know how many different chaotic fields may contribute
to the dark energy of the universe.

Of course, the actual properties
of the chaotic
dark energy component depend on the classical equation of state of
the chaotic fields, which will be dealt with in the next section.

\section{Energy density, pressure, and equation of state}
The kinetic energy term of our chaotic fields is given by
\begin{equation}
T_{kin}=\frac{1}{2} \left(\frac{\partial}{\partial t}
\varphi\right)^2.
\end{equation}
Discretized with lattice constant $\delta$ we obtain for the
expectation of $T_{kin}$
\begin{eqnarray}
\langle T_{kin} \rangle &=& \frac{1}{2}\frac{p_{max}^2}{\delta^2}
\langle (\Phi_n^i -\Phi_n^{i-1})^2 \rangle \nonumber \\ &=&
\frac{p_{max}^2}{\tau}\frac{1}{2} \alpha (\langle \Phi^2 \rangle
-\langle \Phi_n^i\Phi_n^{i-1} \rangle )
\end{eqnarray}
In particular, for $\alpha \to 0$ the expectation of kinetic
energy vanishes, and a universe mainly filled with such a field is
vacuum energy dominated.

In general, the expectation of the total energy density $\langle
\rho \rangle$ is given by
\begin{equation}
\langle \rho \rangle = \langle T_{kin} \rangle +\langle V \rangle
\end{equation}
and the expectation of the pressure by
\begin{equation}
\langle p \rangle =\langle T_{kin} \rangle -\langle V \rangle .
\end{equation}
For the map $T_{-3}$ one obtains
\begin{eqnarray}
\langle \rho \rangle &=&\frac{p_{max}^2}{\tau} \left\{
\frac{\alpha}{2} (\langle \Phi^2 \rangle -\langle
\Phi_n^i\Phi_n^{i-1}\rangle) +(1-\alpha)( \frac{3}{2}\langle
\Phi^2\rangle -\langle \Phi^4 \rangle )\right\}\label{sbarba1} \\
 \langle p \rangle
&=&\frac{p_{max}^2}{\tau} \left\{ \frac{\alpha}{2} (\langle \Phi^2
\rangle -\langle \Phi_n^i\Phi_n^{i-1}\rangle)
-(1-\alpha)(\frac{3}{2}\langle \Phi^2\rangle -\langle \Phi^4
\rangle )\right\},\label{sbarba2}
\end{eqnarray}
where the additive constant of the self-interacting
potential is fixed by the symmetry consideration of section 3. The
above equations
yield the equation of state
\begin{equation}
w=\frac{\langle p\rangle}{\langle \rho \rangle}
\end{equation}
which varies as a function of the coupling $\alpha$ in a
nontrivial way.

For $\alpha =0$, the equation of state of
our fields is $w=-1$, since the expectation of kinetic energy
vanishes (all expectations should be interpreted as quantum
mechanical expectations with respect to second quantization). For
small $\alpha$, $w$ is close to $-1$. It should be clear that although
our fields fluctuate rapidly in both physical and fictitious time,
these fluctuations are averaged away when doing the quantum
mechanical expectations. Thus the classical picture that arises
out of this 2nd quantized rapidly fluctuating
model is a very homogeneous field.

The expectations in eqs.~(\ref{sbarba1}), (\ref{sbarba2}) are easily
numerically calculated by long-term iterating the coupled map
lattice for random initial conditions and averaging over all $i$
and $n$. We used lattices of size 10000 with periodic
boundary conditions. The result for the
equation of state $w(\alpha)$ is displayed in Fig.~1. For small
$\alpha$, $w$ grows
approximately in a linear way. It monotonously increases from
$w=-1$ for $\alpha =0$ to $w=+1$ for $\alpha=1$, up to a wiggle at
$\alpha \approx 0.12$. Fig.~2 shows the corresponding (classical) energy
density and pressure of the field. To account for the currently
observed dark energy in the universe, most
chaotic fields must have a coupling $\alpha$
smaller than about 0.04. Larger
$\alpha$ are ruled out by the observations providing evidence for
$w<-0.78$ \cite{dark}.

\section{Einstein equations and dynamical evolution}
Let us now consider the Einstein equations
for a homogeneous and isotropic universe that consists of three different
components: matter, radiation, and chaotic fields. These
three components are labeled by the indices $m,r,\varphi$, respectively.
One has
\begin{eqnarray}
H^2&=&\left( \frac{\dot{a}}{a} \right)^2=\frac{8}{3}\pi G
(\rho_\varphi +\rho_m +\rho_r)  \\ \frac{\ddot{a}}{a}
&=&-\frac{4}{3}\pi G(\rho_\varphi +3p_\varphi +\rho_m +\rho_r
+3p_r).\label{einstein2}
\end{eqnarray}
Here $\rho_j$ denotes the (classical) energy density of component $j$, and
$p_j$ the pressure. For simpler notation we omit
the expectation values $\langle \cdots \rangle$.
The equation of state of each component is
$w_j=p_j/\rho_j$. As it is well known,
one has for matter $w_m=0$, for radiation
$w_r=1/3$, whereas the equation of state $w_\varphi$ of the
chaotic fields $\varphi$ depends on the coupling $\alpha$ (see
Fig.~1).

The Einstein (or Friedmann)
 equations are usually supplemented by the assumption
of conservation of energy for each species $j$,
\begin{equation}
\dot{\rho_j}=-3H(\rho_j+p_j)\label{econs}.
\end{equation}
These equations can be derived from the Einstein equations under the
assumption of adiabatic expansion, i.e.\ one assumes
that no entropy is produced.

For a universe dominated by a species $j$ with constant equation
of state $w_j=p_j/\rho_j$ eq.~(\ref{econs}) leads to
\begin{equation}
\rho_j\sim a^{-3(1+w_j)} . \label{wj}
\end{equation}
We obtain the well-known result that for matter
$\rho_m\sim a^{-3}$, for radiation $\rho_r\sim a^{-4}$, whereas
for true classical vacuum energy (a cosmological constant
$\Lambda$) with $w=-1$ one has no dependence on $a$ at all,
$\rho_\Lambda =const$.

For the chaotic fields, energy conservation is a non-trivial issue,
for the
following reasons: i) These fields model vacuum fluctuations, and 
vacuum fluctuations by definition do not conserve energy ii) The chaotic
field dynamics (as any chaotic dynamics) constantly produces entropy
in fictitious time, whereas the Friedmann equations (containing no fictitious time)
describe an adiabatic
expansion of the universe iii) The coupling constant
$\alpha$ and hence the equation of state of the chaotic field
may change in time  iv)
There may be an entire spectrum of chaotic fields with different couplings $\alpha$ who can interact with each other v)
The chaotic fields may interact with dark matter.

In the following, we want to restrict ourselves to a simple scenario where
the coupling $\alpha$ is small and where there is energy conservation
of quantum mechanical expectations
in full agreement with
the Friedmann equations. Two interesting possibilities arise in this
context:

a) a {\bf symmetric phase} of the universe, where both the
negative and positive Tchebyscheff dynamics are active. In this
case the vacuum energies $\rho_\varphi^-:=\langle V_{-3}(\varphi)
\rangle$ and $\rho_\varphi^+:=\langle V_{+3}(\varphi) \rangle$
cancel to zero:
\begin{equation}
\rho_\varphi=\rho_\varphi^-+\rho_\varphi^+=0
\end{equation}
(a consequence of eq.~(\ref{symme})). Since the total vacuum energy
$\rho_\varphi$ is zero, there is no statement from the Einstein
(or Friedmann) equations on the time evolution of the absolute
value $|\rho_\varphi^\pm|$. We are free to postulate the validity
of eq.~(\ref{33})
\begin{equation}
|\rho_\varphi^\pm|=|\langle V_{\pm 3}(\varphi) \rangle |\sim
\frac{p_{max}^2}{\tau} \sim \frac{m_{Pl}^2}{t^2} \label{e1}
\end{equation}
for arbitrary couplings $\alpha$.

b) an {\bf asymmetric phase} of the universe, where only the
negative Tcheby\-scheff dynamics $T_{-3}$ is active (or
where it dominates as compared to $T_{+3}$). In this case
one has
\begin{equation}
\rho_\varphi  >0.
\end{equation}
If there is a chaotic field with coupling constant $\alpha$ that
does not interact with other fields then
energy conservation implies
\begin{equation}
\rho_\varphi \sim a^{-3(1+w_\varphi)},
\end{equation}
where $w_\varphi=w_\varphi(\alpha)$ is the equation of state as a
displayed in Fig.~1. 
Note that chaotic fields with $w_\varphi \approx
0$ $(\alpha \approx 0.25)$ behave similar as dark matter, whereas
chaotic fields with $w_\varphi \approx -1$ $(\alpha \approx 0)$ act like a
cosmological constant (see also \cite{padma}).

We now show that eq.~(\ref{e1}), derived from the uncertainty relation,
naturally leads
to tracking behavior of vacuum energy density.
Suppose the universe surrounding the chaotic fields is dominated
by a species with equation of state $w>-1$ (typically matter or
radiation), then
\begin{equation}
a(t)\sim t^{\frac{2}{3(1+w)}} \Longleftrightarrow t^{-2}\sim
a^{-3(1+w)}.
\end{equation}
Putting this into eq.~(\ref{e1}) we obtain
\begin{equation}
|\rho_\varphi^\pm| =|\langle V_{\pm 3}(\varphi)\rangle |\sim
a^{-3(1+w)},\label{rw}
\end{equation}
i.e.\ the vacuum energy density associated with the chaotic fields
decays {\em in the same way} with $a$ as the density of the
dominating species.




Eq.~(\ref{rw}) can help to naturally avoid the cosmological
coincidence problem. Consider e.g.\ the following scenario.
Initially (say, shortly after inflation) we may have a symmetric
phase where $\rho_r \sim |\rho_\varphi^\pm|
>> \rho_m$. Here $\rho_\varphi^-$ denotes the
vacuum energy density of the negative Tchebyscheff map, which is
cancelled by the vacuum energy density
$\rho_\varphi^+=-\rho_\varphi^-$ of the positive Tchebyscheff map.
Then $|\rho_\varphi^\pm|$ first decays approximately as $a^{-4}$,
since the universe is radiation dominated. At some stage we arrive
at $\rho_r\sim |\rho_\varphi^\pm|\sim \rho_m$, and from then on
matter dominates over radiation, so that from then on
$|\rho_\varphi^\pm|$ decays approximately as $a^{-3}$. During the
late-time evolution of the universe, $|\rho_\varphi^\pm|$ will
always stay of the same order of magnitude as $\rho_m$, since both
$|\rho_\varphi^\pm|$ and $\rho_m$ decay as $a^{-3}$. In spite of
this, for small enough couplings $\alpha$ the chaotic fields have
a classical equation of state close to $w_\varphi=-1$, and can
thus produce the accelerated expansion of the universe via
eq.~(\ref{einstein2}), provided there is symmetry breaking between
$T_{+3}$ and $T_{-3}$ at some late stage of the evolution of the
universe. A concrete mechanism for this will be
discussed in section 9.

What is our physical interpretation of the chaotic fields in the
universe? For $\alpha =0$, it can be rigorously proved that
rescaled deterministic chaotic Tchebyscheff maps can be used to
generate spatio-temporal Gaussian white noise on a larger scale
\cite{chaosqft,book}. In other words, on fictitious time scales
$\tau'
>>\tau$ and physical time scales $\delta' >>\delta$ the chaotic noise just
looks like ordinary Gaussian white noise. We may thus couple the
chaotic fields $\varphi$ to ordinary standard model fields in
order to second quantize the standard model fields, i.e.\ use the
chaotic fields as a source of quantization noise. This is the
basic idea of the so-called 'chaotic quantization' approach
\cite{chaosqft}. The chaotic fields are well embedded in this way
and since they are just playing the role of quantization noise, we
do not expect them to have any disturbing influence on, say,
baryogenesis and similar processes in the early universe. In this
interpretation vacuum energy just arises out of the expectation of
a classical potential that generates quantization noise.

\section{Prediction of $\Omega_\varphi$ and $\Omega_m$}
Our approach allows for the prediction of the order of magnitude
of cosmological parameters such as
$\Omega_\varphi=\rho_\varphi/\rho_c$ and $\Omega_m=\rho_m/\rho_c$
at the present time. Let us start from the uncertainty relation in
the form
\begin{equation}
\Delta E\Delta t=\frac{\hbar}{2},
\end{equation}
which implies
\begin{equation}
p_{max}=\frac{1}{2t}.
\end{equation}
Choosing the time scale $\tau =\kappa m_{Pl}^{-2}$ we get for
$\alpha \approx 0$
\begin{equation}
\rho_\varphi^-= \langle V_{-3}(\varphi)\rangle =\frac{3}{8}
\frac{p_{max}^2}{\tau} =\frac{3}{32}\frac{m_{Pl}^2}{\kappa}
\frac{1}{t^2}\label{vex}
\end{equation}
During the radiation dominated period of the universe one has for
the energy density of radiation
\begin{equation}
\rho_r =\frac{\pi^2}{30} N(T) T^4,
\end{equation}
where $T$ is the temperature and $N(T)$ is the number of
relativistic particle degrees of freedom. There is also a relation
between time and temperature, namely
\begin{equation}
t=\sqrt{\frac{90}{32\pi^3N(T)}} \frac{m_{Pl}}{T^2}. \label{tT}
\end{equation}
Putting eq.~(\ref{tT}) into (\ref{vex}) one obtains
\begin{equation}
\rho_\varphi^-=\frac{1}{30}\pi^3\frac{1}{\kappa}N(T)T^4=\frac{\pi}{\kappa}
\rho_r.
 \label{almutly}
\end{equation}
This equation once again shows that it is reasonable to assume
that $\rho_r$ and $\rho_\varphi^-$ have the same order of
magnitude. Since $\rho_\varphi^-$ decays in the same way as
$\rho_r$, eq.~(\ref{almutly}) is valid during the entire radiation
dominated epoch. Finally $\rho_r$ falls below $\rho_m$ and from
then on we have
\begin{equation}
\rho_\varphi^- \approx \frac{\pi}{\kappa}\rho_m. \label{ratio}
\end{equation}
After symmetry breaking we have $\rho_\varphi^-
=\rho_\varphi \approx const$. This implies a prediction for
$\Omega_\varphi:=\rho_\varphi/\rho_c$ at the present time, 
 namely
\begin{equation}
\Omega_\varphi =
\frac{\rho_\varphi}{\rho_\varphi+\rho_m+\rho_r}\approx
\frac{\pi/\kappa}{1+\pi/\kappa }, \label{333}
\end{equation}
neglecting $\rho_r$ at the present epoch and assuming that
the symmetry breaking took place rather recently. In section 4 we saw that
the currently observed dark energy density is best fitted by the
value $\kappa =1.10\pm 0.10$. Eq.~(\ref{333}) yields with this
value the prediction
\begin{equation}
\Omega_\varphi \approx 0.74 \;\;\; \Omega_m\approx 0.26,
\end{equation}
which is consistent with observations\cite{dark}.
Again, due to the reasons that were already
mentioned at the end of section 4,
eq.~(63) should only be regarded as an order-of-magnitude estimate.

\section{Fixing fundamental parameters}

We have seen that chaotic fields can generate the right amount of
vacuum energy
and have the scope to avoid the cosmological
constant and coincidence problem. We now show that they
also offer solutions to the problem of uniqueness of vacua.

First, let us slightly generalize the chaotic field dynamics
(\ref{sym}) to
\begin{equation}
\Phi_{n+1}^i=(1-\alpha)T(\Phi_n^i)+\sigma \frac{\alpha}{2}
(T^b(\Phi_n^{i-1})+T^b(\Phi_n^{i+1})) \label{cs}
\end{equation}
(we neglect the small noise term). The case $\sigma =+1$ is
called 'diffusive coupling', the case $\sigma =-1$ 'anti-diffusive
coupling'. Chaotic fields with $b=1$ are called to be of 'type A' (
$T^1(\Phi)=:T(\Phi))$, chaotic fields with $b=0$ to be of
'type B' ($T^0(\Phi)=:\Phi$).
In \cite{physicad} the chaotic fields were called 'chaotic strings',
but this is only a different name for the same dynamics.
Our derivation in section 2 lead to chaotic fields of B-type with
diffusive coupling, but from a dynamical
systems point of view all 4 degrees of freedom
($b=0,1,\;\sigma=\pm 1$) exist and are of relevance.
As shown in detail in \cite{physicad}, there are two different
types of vacuum energies for the chaotic fields, namely

1. the self energy
\begin{equation}
V(\alpha) := \frac{p_{max}^2}{\tau}\left( \frac{3}{2}\langle \Phi^2 \rangle
-\langle \Phi^4\rangle \right) \label{pose}
\end{equation}
and

2. the interaction energy
\begin{equation}
W(\alpha) := \frac{p_{max}^2}{2\tau} \langle \Phi_n^i \Phi_n^{i+1} \rangle 
\end{equation}
Eq.~(\ref{pose}) actually represents the self energy of the map $T_{-3}$,
the self energy of the map $T_{+3}$ has opposite sign and cancels the
self energy of $T_{-3}$ in the
symmetric phase.

Basically, the self energy is the expectation of the potential
that generates the chaotic dynamics in fictitious time,
and the interaction energy is the expectation of the potential that
generates the diffusive coupling in physical time. 
One may also
define a total vacuum energy as $H^\pm(\alpha):=V(\alpha)\pm \alpha W(\alpha)$,
where the $-$ sign corresponds to diffusive and the $+$ sign to
anti-diffusive coupling. All additive constants
are fixed
by the postulate of invariance of the theory under global and
local $Z_2$-transformations \cite{book}. In other words,
we allow for the existence of a symmetric phase.
For small $\alpha$ the interaction energy
can be neglected as compared to the self energy, moreover,
the type-A and type-B forms are observed to have the same self energy
in this limit.

The central hypothesis of this paper is a symmetry between
standard model coupling constants and the chaotic field
couplings $\alpha$. We assume that for
any dimensionless coupling constant $\alpha$ that appears in the
standard model of electroweak and strong interactions, there is a
corresponding chaotic field that is just coupled with this
$\alpha$. The universe then
tries to reach a state of minimum
vacuum energy by adjusting its free parameters in such a way
that the chaotic fields reach a state of minimum vacuum energy.

While at first sight this may look like a purely theoretical concept, there is
numerical
evidence that this principle is indeed physically realized. As an example,
Fig.~3 shows the self energy
$V(\alpha)=\langle V_{-3}(\varphi)\rangle$ of our chaotic fields
of type A with diffusive coupling
in the low-coupling region. We observe that $V(\alpha)$ has
local minima at
\begin{eqnarray}
a_{1} &=& 0.000246(2) \\ a_{2} &=& 0.00102(1)
\\ a_{3} &=& 0.00220(1)
\end{eqnarray}
($a_1$ and $a_3$ are actually small local minima
on top of the hill).

On the other hand, in the standard model of electroweak
interactions the weak coupling constant is given by
\begin{equation}
\alpha_{weak} = \alpha_{el} \frac{(T_3-Q\sin^2 \theta_W)^2}{\sin^2
\theta_W \cos^2\theta_W} \label{aweak}
\end{equation}
Here $Q$ is the electric charge of the particle ($Q=-1$ for
electrons, $Q=2/3$ for $u$-like quarks, $Q=-1/3$ for $d$-like
quarks), and $T_3$ is the third component of the isospin ($T_3=0$
for right-handed particles, $T_3=-\frac{1}{2}$ for $e_L$ and
$d_L$, $T_3=+\frac{1}{2}$ for $\nu_L$ and $u_L$). Consider
right-handed fermions $f_R$. With $\sin^2 \theta_W =\bar{s}_l^2=0.2318$ (as
experimentally measured \cite{pada}) and the running electric coupling
$\alpha_{el} (E)$ taken at energy scale $E=3 m_f$ we obtain from
eq.~(\ref{aweak}) the
numerical values
\begin{eqnarray}
\alpha_{weak}^{d_R} (3 m_d) &=&0.000246 \\ \alpha_{weak}^{c_R} (3
m_c) &=&0.001013 \\ \alpha_{weak}^{e_R} (3 m_e) &=&0.00220 .
\end{eqnarray}
There is an amazing numerical coincidence between the local minima
$a_1,a_2,a_3$ of $V(\alpha)$ and the weak coupling constants of
$f_R=u_R,c_R,e_R$, respectively.

Now regard the fine structure constant $\alpha_{el}$ and the
Weinberg angle $\sin^2 \theta_W$ as a priori free parameters.
Suppose these parameters would change to slightly different
values. Then immediately this would produce larger vacuum energy
$V(\alpha)$, since we get out of the local minima. The system is
expected to be driven back to the local minima, and the
fundamental parameters are stabilized in this way, provided the universe
is in an asymmetric phase.


The above example is only one example of a large number of
numerical coincidences
observed. In \cite{book,physicad} an extensive numerical
investigation of self energies, interaction
energies, and total vacuum energies was performed for the above
chaotic field theories. A large number of amazing numerical
coincidences was found\footnote{For the $\varphi^4$-theory
the relevant energy
scale is always $E=3m_f$. The factor 3 can be related
to the index of the Tchebyscheff polynomial considered \cite{book}.}.
These results are described in detail in
\cite{book,physicad}, we here only summarize the main results.

\begin{enumerate}
\item The smallest (stable) zeros of the interaction energy
$W(\alpha)$ coincide with running electroweak coupling constants,
evaluated at energies given by the smallest fermionic mass scales.
Type (A) describes $d$-quarks and electrons interacting
electrically, type (B) $u$-quarks
and neutrinos interacting weakly.

\item Local minima of the self energy $V(\alpha)$ coincide with
running weak coupling constants of right-handed fermions,
evaluated at the lightest fermionic mass scales.

\item Local minima of the total vacuum energy $H^+(\alpha)$
occur at running strong coupling constants evaluated at the
lightest baryonic energy scales.

\item Local minima of the total vacuum energy $H^-(\alpha)$
occur at running strong couplings evaluated at the lightest
mesonic energy scales.
\end{enumerate}

In \cite{book,physicad} also chaotic fields corresponding to 2nd
order Tchebyscheff maps were investigated (see Appendix), and the
following numerical coincidences were found:

\begin{enumerate}
\item The smallest (stable) zeros of $W(\alpha)$ coincide
with running strong coupling constants evaluated at the smallest
bosonic mass scales. Type (A) describes the $W$ boson, type (B)
the Higgs boson.
\item Local minima of the self energy $V(\alpha)$ coincide with
Yukawa and gravitational couplings evaluated at the fermionic mass
scales.
\end{enumerate}
For more details, see \cite{book,physicad}.

All these numerically observed coincidences are not explainable
as a random coincidence. Rather, they suggest to interpret
the coupling constant $\alpha$ of our second-quantized chaotic fields
$\varphi$ as a running gauge coupling. We are free to identify
$\alpha=2\tau/\delta^2$ with a gauge coupling, since the occurence
of a ratio of lattice constants $\tau$ and $\delta^2$ is a new
effect in our 2nd quantized discretized theory, and there is no
theory of this dimensionless number so far,
which represents a kind of metric for the 5th coordinate
(the fictitious time).  So we are indeed free
to make the hypothesis that $\alpha$ coincides with a running
gauge coupling. By doing so, we implicitly construct a symmetry
between gauge couplings and gravitational couplings, since usually
the strength of the kinetic term in the action of a field is
determined by the metric, i.e.\ gravitational effects, whereas
here it is fixed by standard model coupling strengths.
The chaotic fields appear to select out of the infinitely many
vacua allowed by string theory the unique ground state that
corresponds to the known coupling constants of the universe.
All free parameters are fixed in the sense that if
the fundamental parameters (masses, coupling constants, and mixing
angles) had different values, larger vacuum energy would
arise.

\section{Spontaneous symmetry breaking and
cancellation of unwanted vacuum energy}

Chaotic scalar fields not only allow for a simple mechanism to produce
dark energy, they also yield a simple mechanism to cancel
unwanted dark energy.
If
we assume that both the positive and negative
Tchebyscheff dynamics are
physically realized, the corresponding vacuum energies precisely cancel
for symmetry reasons (see
eq.~(\ref{symmetry})). This symmetry is a $Z_2$ symmetry which is not
there for ordinary smoothly evolving
scalar fields (where opposite potentials lead to unstable
or ill-defined behavior).
On the other hand,
if only the negative Tchebyscheff field dynamics is active,
or if it dominates,
then positive
dark energy arises. This positive dark energy can
drive inflation, fix standard model
parameters as local minima in the dark energy
landscape, and generate
late-time acceleration. It is therefore desirable to construct a theory
that allows for $Z_2$ symmetry breaking between positive and negative
Tchebyscheff maps.

It is clear that in order to fix fundamental parameters
with the methods described in the previous section, we must
have a broken $Z_2$ symmetry at some stage of the
evolution of the universe. Indeed, the minimum
requirement we need is
at least one very early stage of broken symmetry, in order
to {\em first-time fix} the fundamental parameters to the values
which make the universe work, and another
late-time asymptotic state of broken
symmetry, in order to {\em stabilize} the parameters to
their known values so that they cannot
drift away to other values. It is natural
to identify the first phase of broken
symmetry with the
inflationary phase \cite{infla}, and the other phase of broken symmetry
with the late-time
state of the universe. Inbetween, we may allow for a symmetric state,
which has the advantage that nucleosynthesis is not
spoilt\footnote{The abundance of light elements
is correctly predicted by standard
big bang nucleosynthesis but is spoilt if there is too much
dark energy \cite{nucleo}.
The measured
cosmic microwave background also seems to indicate
little or no dark energy at the time of last
scattering \cite{cmb}.
Galaxy formation is disturbed as well if there is too much
dark energy \cite{galaxy}.}.

As a concrete simple model, consider a scalar field $\sigma$ which takes
on the value $\sigma = 0$ in the symmetric phase
and the values
$\sigma =\pm 1$ in the phase where the $Z_2$-symmetry is spontaneously
broken. The total potential describing the chaotic field dynamics
is given by
\begin{equation}
V (\sigma , \Phi_-, \Phi_+)=\frac{1-\sigma}{2}
V_{-3}(\Phi_-)+\frac{1+\sigma}{2}V_{+3}(\Phi_+), \label{totalv}
\end{equation}
where
\begin{equation}
V_{-3}(\Phi_-) =\frac{p_{max}^2}{\tau} (\Phi_-^2-\Phi_-^4+
\frac{1}{2} \langle \Phi_-^2 \rangle )
\end{equation}
is the potential generating the negative Tchebyscheff field dynamics
and
\begin{equation}
V_{+3}(\Phi_+) =\frac{p_{max}^2}{\tau} (-2\Phi_+^2
+\Phi_+^4+\frac{1}{2} \langle \Phi_+^2 \rangle )
\end{equation}
the one generating the positive Tchebyscheff field dynamics
in fictitious time ($\alpha \approx 0$). In the symmetric phase
($\sigma =0$) we obtain from eq.~(\ref{totalv})
\begin{equation}
\langle V(\sigma ,\Phi_-, \Phi_+) \rangle =0,
\end{equation}
wheras in a broken phase with $\sigma =-1$ we obtain
\begin{equation}
\langle V (\sigma, \Phi_-, \Phi_+) \rangle =\langle V_{-3} (\Phi_- )\rangle
=\frac{p_{max}^2}{\tau} (\frac{3}{2} \langle \Phi^2 \rangle
-\langle \Phi^4 \rangle ) >0, \label{posvac}
\end{equation}
where we have re-labeled $\Phi_-=\Phi$.

We assume that the symmetry
is first spontaneously broken to $\sigma =-1$ at the onset of inflation.
A large amount of positive vacuum energy is generated via
eq.~(\ref{posvac}),
since at this stage the universe is very young
and $p_{max}\sim t^{-1}\sim H$.
The chaotic fields can help to drive inflation,
and fundamental parameters are pre-fixed as local minima in
the dark energy landscape.

Then, there is a symmetric phase with $\sigma =0$.
The consideration of section 7 applies
but the dark energy is suppressed due to symmetry reasons.
Big bang nucleosynthesis and galaxy formation can
go ahead without any problems.
Note that during the symmetric epoch
the fundamental parameters are no longer stabilized as local
minima in the dark energy landscape.
They can drift to slightly different values.
This is consistent
with the experimental findings of a
varying fine structure constant \cite{webb}.

Finally, there is late-time symmetry breaking to $\sigma =-1$.
This phase is necessary because otherwise the
fundamental parameters would keep on drifting to
different values. By the late-time symmetry breaking,
the parameters are finally forced back and stabilized
at their equilibrium values, which were already pre-fixed
during inflation.
This gives physical sense to the
role of late-time dark energy in the universe.

Late-time symmetry breaking could be physically understood as follows.
Suppose
the
negative Tschebyscheff dynamics (which generates positive
vacuum energy) is always uniformly distributed in space. It is a
property of empty space-time.
At a very
early stage, the corresponding vacuum energy was strong and
may have driven inflation.
Then matter and radiation is created. Assume
that the matter and radiation particles
are second-quantized by chaotic noise generated by the positive
Tchebyscheff map (which generates negative vacuum energy). Then,
as soon as sufficiently many
matter and radiation have been created and quantized, the two vacuum energies
$\langle V_{+3}\rangle$ and $\langle V_{-3}\rangle$ 
can precisely (or almost) cancel, 
as long as matter and radiation
are uniformly distributed in space. Inflation
may stop in this way and we obtain a symmetric phase of the 
universe after inflation.
But at a late stage of the evolution of the universe 
matter clumps into galaxies. 
Since the negative vacuum energy is generated by chaotic quantization
noise for each particle it follows the spatial distribution of matter.
The positive vacuum energy remains uniformly distributed.
Hence after structure formation 
there is no spatially uniform cancellation of
vacuum energy anymore. Empty space has an excess of positive
vacuum energy, galaxies are spatial regions with an excess of negative
vacuum energy due to quantization noise (the negative vacuum energy
in the galaxy can be partially compensated by positive
kinetic terms that arise out of spatial inhomogenities in the galaxy).
In this physical interpretation the
late-time symmetry breaking is related to structure formation.

\section{Conclusion}

We have presented a new model for dark
energy in the universe. This model is based on a rather
conservative approach, the assumption of the existence of second
quantized self-interacting scalar fields described by a
$\varphi^4$-theory. However, the main difference is that these
fields are strongly self-interacting, rather than weakly. When
doing 2nd quantization using the Parisi-Wu approach, rapidly
fluctuating chaotic fields arise. The expectation of the
underlying potentials yields the currently observed dark energy
density.

The advantage of this new chaotic model is that many of the
questions raised in the introduction seem to have natural
solutions. The cosmological constant problem is avoided,
in our model the right order of magnitude of
vacuum energy is naturally produced if we interpret the chaotic
dynamics in terms of vacuum fluctuations allowed by the
uncertainty relation, for a given finite age of the universe.
The cosmological coincidence problem is also
avoided, since in our model
the generated dark energy
is not constant anymore, but thins out with the expansion of the
universe in the same way as the energy density of the dominating
species (matter or radiation). In spite of that, the 
(classical)
equation of state of the chaotic component is close to
$w=-1$, and can account for the accelerated expansion of the universe,
provided there is late-time symmetry breaking.
The chaotic fields are physically interpreted in terms of
vacuum fluctuations.
As such they can temporarily
violate energy conservation, 
but quantum mechanical expectations
are fully compatible with the
Friedmann equations.

The physical relevance of our model is emphasized by the
observation of a large number of numerical coincidences between
local minima in the dark energy landscape and running standard
model coupling constants evaluated at the known
fermionic and bosonic mass scales. It
thus appears that chaotic fields have the potential to fix
and stabilize
fundamental parameters and to select the physically
relevant vacuum state out of infinitely many possibilities.

\section*{Appendix A: General Tchebyscheff maps}

Our approach can be easily generalized to Tchebyscheff maps of
arbitrary order $N$. One has $T_1(\Phi) =\Phi$,
$T_2(\Phi)=2\Phi^2-1$, $T_3(\Phi)=4\Phi^3-3\Phi$, generally
$T_N(\Phi)=\cos (N\arccos \Phi )$ with $\Phi \in [-1,1]$. A
Tchebyscheff map of order $N$ is conjugated to a Bernoulli shift
of $N$ symbols, it is ergodic and mixing for $N\geq 2$. It
exhibits the strongest possible chaotic behaviour that is possible
for a 1-d smooth map, characterized by a minimum sceleton of
higher-order correlations \cite{beckhilgers}.

It is useful to consider both positive and negative Tchebyscheff
maps and to define
\begin{equation}
T_{-N} (\Phi):=-T_N(\Phi).
\end{equation}
The behaviour of $T_{-N}$ under iteration is identical to that of
$T_N$ up to a sign, the trajectory of $T_{-N}$ differs by a
constant sign ($N$ even) or an alternating sign ($N$ odd) from
that of $T_N$.

Eq.~(\ref{pot}) implies that the maps $T_N$ correspond to
potentials $V_N$ given by
\begin{equation}
V_N(\varphi) =\frac{1-\alpha}{\tau} \left\{ -\frac{1}{2} \varphi^2 +
p_{max}\int d\varphi \,T_N\left(\frac{\varphi}{p_{max}}\right)
\right\} + const
\end{equation}
In particular, one obtains for $N=\pm 1,\pm 2,\pm 3$
\begin{eqnarray}
V_{\pm 1}(\varphi)&=&\frac{1-\alpha}{\tau}\left( -\frac{1}{2} \varphi^2
\pm \frac{1}{2} \varphi^2 \right)+ const
\\ V_{\pm 2}(\varphi)&=& \frac{1-\alpha}{\tau} \left( -\frac{1}{2}
\varphi^2\pm (\frac{2}{3p_{max}}\varphi^3-p_{max}\varphi) \right)
+const
\\ V_{\pm 3}(\varphi) &=& \frac{1-\alpha}{\tau} \left(-\frac{1}{2}
\varphi^2 -\pm ( -\frac{3}{2} \varphi^2
+\frac{1}{p_{max}^2}\varphi^4 )\right) +const .
\end{eqnarray}
Of physical relevance are the expectations of these potentials,
formed with respect to the ergodic
dynamics. Since negative and positive Tchebyscheff maps generate
essentially the same dynamics, up to a sign, any physically
relevant expecatation should also be the same for $T_N$ and
$T_{-N}$, up to a possible sign. For all $N$, this symmetry
condition fixes the additive constant to be
\begin{equation}
const=+\frac{1-\alpha}{\tau} \frac{1}{2}\langle \varphi^2 \rangle
\end{equation}
With this choice one obtains the following formulas for the self
energy which are fully symmetric under the transformation $N\to
-N$:
\begin{eqnarray}
\langle V_{\pm 1} (\varphi)\rangle &=&\pm \frac{1-\alpha}{\tau}
\frac{1}{2} \langle \varphi^2 \rangle \\ \langle V_{\pm 2}
(\varphi) \rangle &=& \frac{1-\alpha}{\tau} \left( \frac{2}{3p_{max}}
\langle \varphi^3 \rangle-p_{max}\langle \varphi \rangle \right)
\\ \langle V_{\pm 3}(\varphi) \rangle &=&\pm \frac{1-\alpha}{\tau} \left(
-\frac{3}{2}\langle \varphi^2 \rangle + \frac{1}{p_{max}^2}\langle
\varphi^4 \rangle \right)
\end{eqnarray}
Written in terms of the dimensionless field variable $\Phi =
\varphi /p_{max}$ this is
\begin{eqnarray}
\langle V_{\pm 1} (\varphi) \rangle &=&\pm
\frac{1-\alpha}{\tau}p_{max}^2 \frac{1}{2} \langle \Phi^2 \rangle \\
\langle V_{\pm 2}(\varphi) \rangle &=& \frac{1-\alpha}{\tau}p_{max}^2
\left( \frac{2}{3} \langle \Phi^3 \rangle- \langle \Phi \rangle
\right)
\\ \langle V_{\pm 3} (\varphi) \rangle &=&\pm \frac{1-\alpha}{\tau}
p_{max}^2
 \left(
-\frac{3}{2}\langle \Phi^2 \rangle + \langle \Phi^4 \rangle
\right).
\end{eqnarray}
For Tchebyscheff
maps of arbitrary order $N$ one obtains
\begin{eqnarray}
V_{\pm N}(\varphi)&=& \frac{1-\alpha}{\tau} p_{max}^2 \left\{
-\frac{1}{2} \Phi^2\pm \int \cos (N \arccos \Phi ) d\Phi \right\}
\\ &=& \frac{1-\alpha}{2\tau}p_{max}^2\left\{
-\Phi^2\pm \left(\frac{1}{N+1}T_{N+1} (\Phi
)-\frac{1}{N-1}T_{N-1}(\Phi )\right) \right\}\nonumber \\ &\,&
 +const
\end{eqnarray}
and
\begin{equation}
\langle V_{\pm N}(\varphi)\rangle = (\pm 1)^N
\frac{1-a}{2\tau}p_{max}^2\left\{ \frac{1}{N+1}\langle T_{N+1}
(\Phi )\rangle -\frac{1}{N-1}\langle T_{N-1}(\Phi )\rangle
+C \right\}.
\end{equation}

For uncoupled Tchebyscheff maps with $|N|\geq 2$, any expectation
of an observable $A (\Phi)$ is given by eq.~(\ref{obs}) and
(\ref{mu}).
For $\alpha \not=0$ the invariant density changes in a nontrivial way,
but expectations can still be easily calculated numerically by
long-time iteration of the coupled map lattice.




\subsection*{Acknowledgement}
I am very grateful to Dr. E. Komatsu for useful discussions.

This research was supported in part by the National Science Foundation
under Grant No. PHY99-07949.

\newpage

\epsfig{file=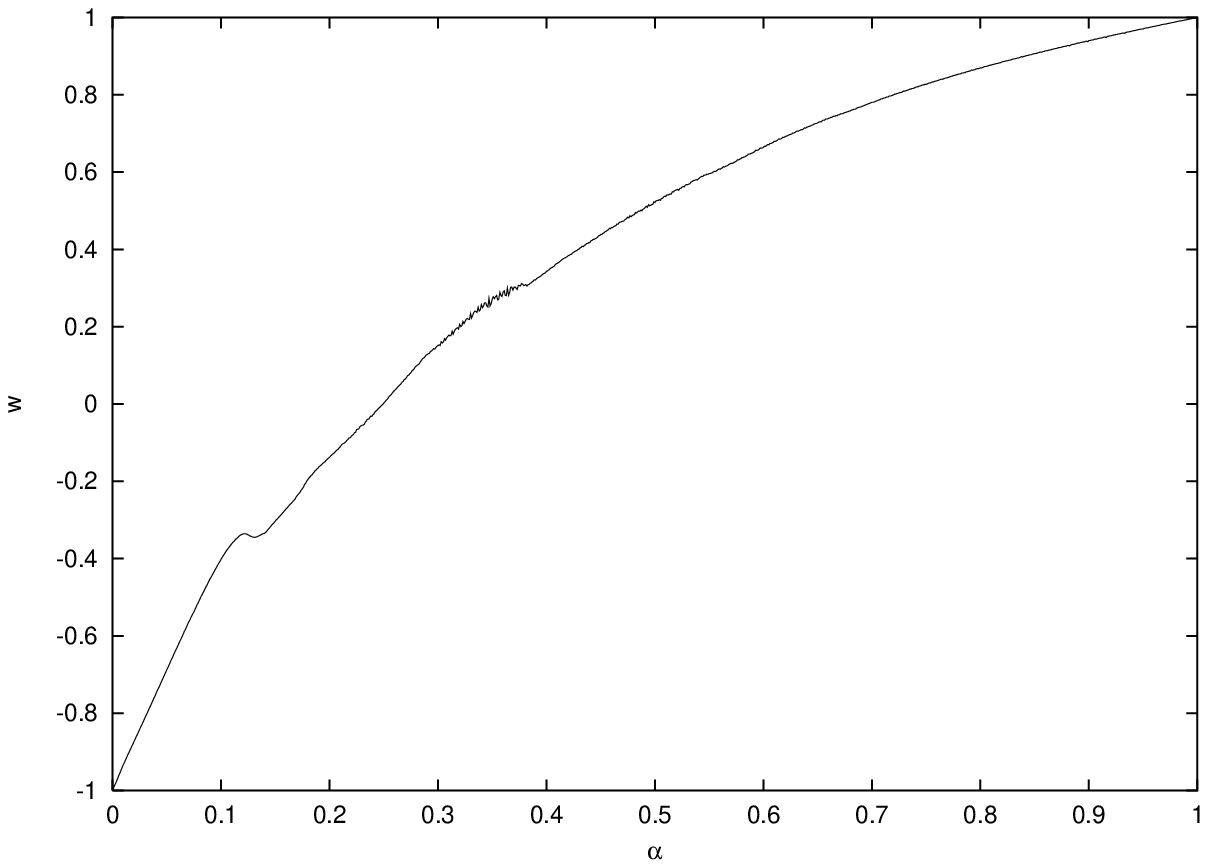}

\vspace{0.5cm}

{\bf Fig.~1} Classical equation of state $w=\langle p \rangle /
\langle \rho \rangle$ of the chaotic field $\varphi$
as a function of the coupling $\alpha$.

\vspace{1cm}

\epsfig{file=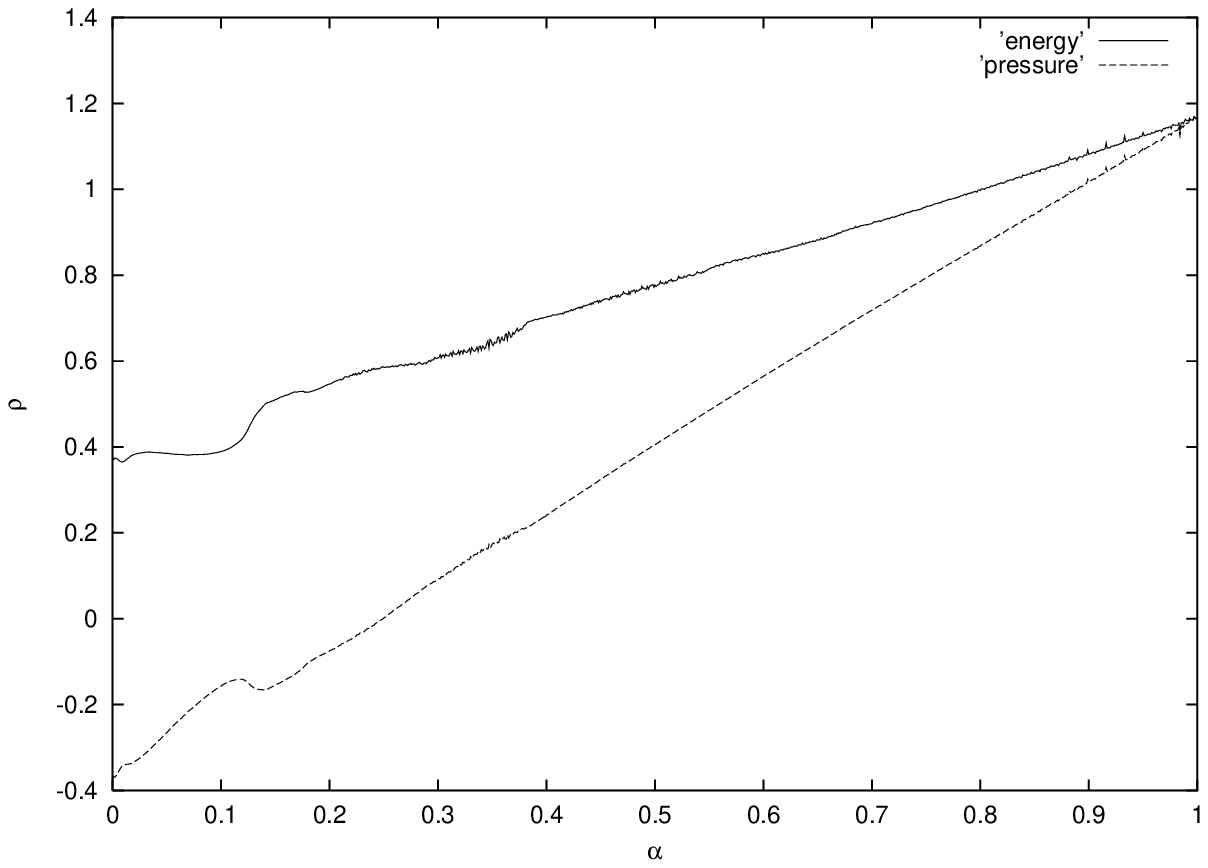}

\vspace{0.5cm}

{\bf Fig.~2} Expectation of energy and pressure of the chaotic field as
a function of the coupling $\alpha$.

\vspace{1cm}

\epsfig{file=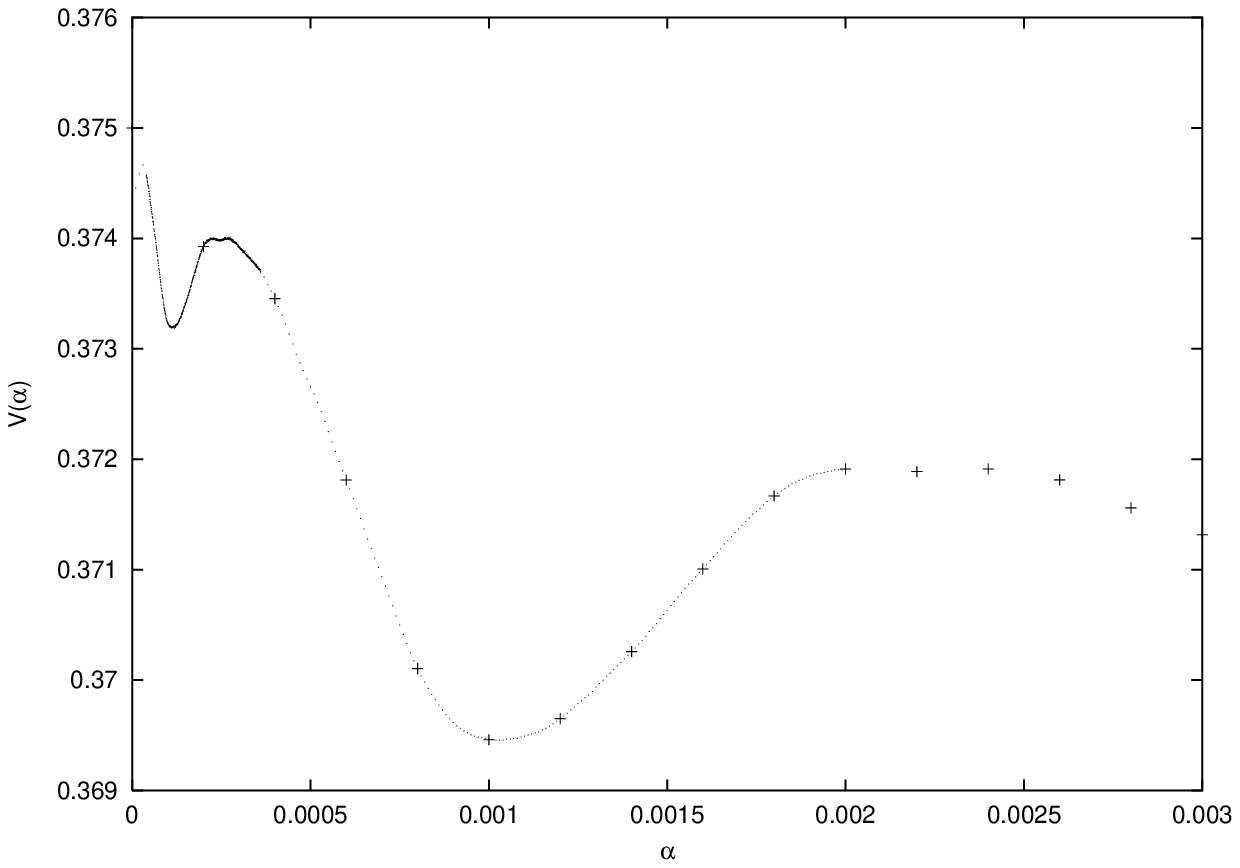}

\vspace{0.5cm}

{\bf Fig.~3} Self energy $V(\alpha)$ of the type-A chaotic field in the
low-coupling region. There are local minima at couplings $a_i$
that coincide with the weak coupling
constants of right-handed fermions in the standard model.

\begin{thebibliography}{99}
\bibitem{accel} A.G. Riess et al., Astron. J. {\bf 116},
1009 (1998), S. Perlmutter et al., Astrophys. J. {\bf 517}, 565 (1999),
J.L. Tonry et al., astro-ph/0305008
\bibitem{accel2} N.W. Halverson et al., Astrophys. J. {\bf 568}, 38 (2002),
C.B. Netterfield et al., Astrophys. J. {\bf 571}, 604 (2002),
\bibitem{dark} D.N. Spergel et al., Astrophys. J. Suppl.
{\bf 148}, 175 (2003) (astro-ph/0302209),
C.L. Bennett et al., Astrophys. J.
Suppl. {\bf 148}, 1 (2003)
(astro-ph/0302207), G. Brumfiel, Nature {\bf 422}, 108 (2003)
\bibitem{cos} S. Weinberg, Rev. Mod. Phys. {\bf 61}, 1 (1989),
S. Caroll, astro-ph/0004075, S. Weinberg, astro-ph/0005265,
U. Ellwanger, hep-ph/0203252, T. Padmanabhan, Phys. Rep. {\bf 380}, 235
(2003) (hep-th/0212290)
\bibitem{quin} P.J.E. Peebles and B. Ratra,
Astrophys. J. {\bf 325}, L17 (1988), C. Wetterich, Nucl. Phys. {\bf B312},
668 (1988),
M.S. Turner and M. White, Phys. Rev. {\bf D56},
4439 (1997),  J.A. Frieman and I. Waga, Phys. Rev. {\bf D57}, 4642 (1998),
R.R. Chadwell, R. Dave, and P.J. Steinhardt, Phys. Rev. Lett. {\bf 80}, 1562 (1998),
P.J.E. Peebles and A. Vilenkin,
Phys. Rev. {\bf 59D}, 063505 (1999)
\bibitem{string} M. Gasperini, F. Piazza, and G. Veneziano, Phys. Rev.
{\bf D65},
023508 (2002), S. Dimopoulous and S. Thomas, hep-th/0307004,
R.H. Brandenberger, hep-th/0103156
\bibitem{phantom} R.R. Caldwell, Phys. Lett. {\bf B545}, 23 (2002),
S.M. Caroll, M. Hoffmann, and M. Trodden, astro-ph/0301273,
J. Hao and X Li, hep-th/0306033
\bibitem{BI} J. Polchinski, {\em String Theory}, Cambridge University Press,
Cambridge (1998), A.A. Tseytlin, hep-th/9908105,
E. Elizalde, J.E. Lidsey, S. Nojiri, and S.D. Odintsov, hep-th/0307177
\bibitem{bafi} T. Banks, W. Fischler, astro-ph/0307459,
T. Banks, hep-th/0007146,
R. Bousso, J. Polchinski, JHEP {\bf 0006}, 006 (2000),
S. Mukohyama and L. Randall,
hep-th/0306108
\bibitem{stoch} G. Parisi, Y. Wu, Sci Sin. {\bf 24}, 483 (1981),
P.H. Damgaard, H. H\"uffel (eds.),
{\em Stochastic Quantization}, World Scientific, Singapore (1988)
\bibitem{chaosgr}
N.J. Cornish, J. Levin, Class. Quant. Grav. {\bf 20}, 2649 (2003),
X. Wu, T. Huang, Phys. Lett. {\bf A313}, 77 (2003),
R. Easther and K. Maeda,
Class. Quant. Grav. {\bf 16}, 1637 (1999),
R.O. Ramos, Phys. Rev. {\bf D64}, 123510 (2001)
\bibitem{chaosqft}
T.S. Bir\'{o}, S.G. Matinyan, B. M\"uller, {\em
Chaos and Gauge Field Theory}, World Scientific, Singapore (1994)
C. Beck, Nonlinearity {\bf 8}, 423 (1995),
R.O. Ramos and F.A.R. Navarro, Phys. Rev. {\bf D62}, 085016 (2000),
T.S. Bir\'{o}, B. M\"uller, S.G. Matinyan,
hep-th/0301131, C. Beck, hep-th/0305173
\bibitem{book} C. Beck, {\em Spatio-temporal Chaos and Vacuum Fluctuations
of Quantized Fields}, World Scientific, Singapore (2002) (50-page
summary at hep-th/0207081)
\bibitem{physicad} C. Beck, Physica {\bf 171D}, 72 (2002)
\bibitem{chaosstring}
T. Damour, M. Henneaux, B. Julia, and H. Nicolai,
Phys. Lett. {\bf B509}, 323 (2001), I. Kogan and D. Polyakov,
hep-th/0212137, B.L. Julia, hep-th/0209170
\bibitem{CML} K. Kaneko, Progr. Theor. Phys. {\bf 72}, 480 (1984),
R. Kapral, Phys. Rev. {\bf A31}, 3868 (1985), C. Beck, Phys. Lett.
{\bf
248A}, 386 (1998), C.P. Dettmann, Physica {\bf 172D}, 88 (2002)
\bibitem{escort}
C. Beck and F. Schl\"ogl, {\em Thermodynamics of Chaotic Systems},
Cambridge University Press, Cambridge (1993)
\bibitem{padma} T. Padmanabhan, T. Roy Choudhury,
Phys. Rev. {\bf D66}, 081301 (hep-th/0205055)
\bibitem{pada} K. Hagiwara et al., Phys. Rev. {\bf D66}, 010001 (2002)
\bibitem{beckhilgers}
C. Beck, Nonlinearity {\bf 4}, 1131 (1991), A. Hilgers, C. Beck,
Physica {\bf 156 D}, 1 (2001)
\bibitem{infla} A. Guth, Phys. Rev. {\bf 23D}, 347 (1981),
A.D. Linde, {\em Particle Physics and Inflationary Cosmology},
Harwood, Chur (1990)
\bibitem{nucleo} M. Yahiro et al., Phys. Rev. {\bf 65D},
063502 (2002) (astro-ph/0106349),
J.P. Kneller and G. Steigman, Phys. Rev. {\bf 67D}, 063501 (2003)
(astro-ph/0210500)
\bibitem{cmb} R.R. Caldwell et al., astro-ph/0302505,
E. Komatsu, private communication
\bibitem{galaxy} S. Weinberg, Phys. Rev. Lett. {\bf 59}, 2607 (1987)
\bibitem{webb} J.K. Webb et al., Phys. Rev. {\bf 82}, 884 (1999),
{\bf 87}, 091301 (2001)
\end{thebibliography}
\end{document}